\documentclass[10pt,conference]{IEEEtran}
\IEEEoverridecommandlockouts

\usepackage{cite}
\usepackage{amsmath,amssymb,amsfonts}
\usepackage{algorithmic}
\usepackage{graphicx}
\usepackage{textcomp}
\usepackage{xcolor}
\usepackage[hidelinks]{hyperref}
\usepackage{tabularx}
\usepackage{booktabs}
\usepackage{comment}
\usepackage{enumitem}
\usepackage[linesnumbered,ruled,vlined,noend]{algorithm2e}
\usepackage[skins]{tcolorbox}
\usepackage{subfigure}
\usepackage{listings}
\usepackage{multirow}
\usepackage{array}

\def\BibTeX{{\rm B\kern-.05em{\sc i\kern-.025em b}\kern-.08em
    T\kern-.1667em\lower.7ex\hbox{E}\kern-.125emX}}

\lstnewenvironment{Report}[1][]{
  \lstset{
    basicstyle =\fontsize{8}{8}\selectfont\ttfamily,
    frame      = tb,
    mathescape = true
  }
}{}

\newtheorem{defn}{Definition} 

\SetKwRepeat{Do}{do}{while}

\makeatletter
\patchcmd\algocf@Vline{\vrule}{\vrule \kern-0.4pt}{}{}
\patchcmd\algocf@Vsline{\vrule}{\vrule \kern-0.4pt}{}{}
\makeatother

\SetCommentSty{mycommfont}
\SetKwComment{Comment}{\ \#\ }{}
\SetKwProg{Func}{function}{}{end}

\newcommand{\Code}[1]{\begin{small}\fontsize{9.5}{10}\selectfont\texttt{#1}\end{small}}

\newcommand{\ToolName}{\textsc{Silence}}

\tcbset{
  my box2/.style={
    enhanced,
    colframe=#1!80,
    colback=#1!5,
  },
}
\newtcolorbox{summary-rq}{
  my box2=black,
  boxrule=1pt,top=3pt,bottom=3pt,left=4pt,right=4pt
}
    
\begin{document}

\title{Understanding and Remediating Open-Source License Incompatibilities in the PyPI Ecosystem}

\author{\IEEEauthorblockN{Weiwei Xu\thanks{\IEEEauthorrefmark{1}Both authors contributed equally to this paper.}\IEEEauthorrefmark{1}, Hao He\IEEEauthorrefmark{1}, Kai Gao, Minghui Zhou\thanks{\IEEEauthorrefmark{2}Minghui Zhou is the corresponding author.}\IEEEauthorrefmark{2}}
\IEEEauthorblockA{\textit{School of Computer Science and School of Software \& Microelectronics, Peking University, Beijing, China}\\
\textit{Key Laboratory of High Confidence Software Technologies, Ministry of Education, China}\\
xuww@stu.pku.edu.cn, \{heh, gaokai19, zhmh\}@pku.edu.cn}
}

\maketitle

\thispagestyle{plain}
\pagestyle{plain}

\begin{abstract}
The reuse and distribution of open-source software must be in compliance with its accompanying open-source license.
In modern packaging ecosystems, maintaining such compliance is challenging because a package may have a complex multi-layered \textit{dependency graph} with many packages, any of which may have an incompatible license.
Although prior research finds that license incompatibilities are prevalent, empirical evidence is still scarce in some modern packaging ecosystems (e.g., PyPI).
It also remains unclear how developers remediate the license incompatibilities \textit{in the dependency graphs} of their packages (including direct and transitive dependencies), let alone any automated approaches.

To bridge this gap, we conduct a large-scale empirical study of license incompatibilities and their remediation practices in the PyPI ecosystem.
We find that 7.27\% of the PyPI package releases have license incompatibilities and 61.3\% of them are caused by transitive dependencies, causing challenges in their remediation;
for remediation, developers can apply one of the five strategies: migration, removal, pinning versions, changing their own licenses, and negotiation.
Inspired by our findings, we propose \ToolName, an SMT-solver-based approach to recommend license incompatibility remediations with minimal costs in package dependency graph.
Our evaluation shows that the remediations proposed by \ToolName~can match 19 historical real-world cases (except for migrations not covered by an existing knowledge base) and have been accepted by five popular PyPI packages whose developers were previously unaware of their license incompatibilities.
\end{abstract}


\section{Introduction}

Open-source licenses dictate the terms and conditions regarding how a piece of open-source software (OSS) can be reused, modified, and redistributed~\cite{DBLP:journals/tosem/XuGFLLJ23}. 
As of April 2023, the Open Source Initiative (OSI) has approved 117 open-source licenses~\cite{OSIApproved}, ranging from highly restrictive ones (e.g., GPL 3.0~\cite{GPLv3}) to highly permissive ones (e.g., MIT~\cite{MIT}).
When developers incorporate OSS into their projects, it is critical to comply with all the terms and conditions declared in the license of the OSS.
Failure to do so can result in ethical, legal, and monetary consequences~\cite{rosen2005open, DBLP:journals/jmis/SojerAKH14}. 

As OSS thrives, modern software development is increasingly dependent on the reuse of OSS packages from major packaging ecosystems (e.g., PyPI~\cite{PyPI}, Maven~\cite{Maven}, npm~\cite{npm}).
On the other hand, the legal risks of reusing OSS packages from packaging ecosystems are high because packages form complex dependency networks in which one package can directly or transitively depend on hundreds of other packages~\cite{DBLP:journals/ese/DecanMG19}. 
Any of the dependent packages may have a very restrictive license, which can easily introduce license violations for any package or downstream project depending on them.

In this paper, we consider the \textit{license incompatibility} issue occurring when an OSS package release\footnote{
In this paper, we align with the terminology of PyPA~\cite{packaging-glossary} and use the term \textit{release} to refer to a specific version of a package.
For example, \texttt{fiftyone 0.18.0} is one of the releases of \texttt{fiftyone} with version number \texttt{0.18.0}.
} 
depends on another release whose license is incompatible with its own license. 
License incompatibilities can arise from both direct and transitive dependencies in a release's dependency graph~\cite{DBLP:journals/jip/QiuGI21, DBLP:conf/icsr/MakariZR22}.
By \textit{dependency graph} (sometimes also referred to as dependency tree~\cite{DBLP:conf/icse/LiuCF00022}), we mean a directed graph with a release as the root node, releases that the root node directly or transitively depends on as other nodes, and direct dependency relationships between nodes as edges.
A dependency graph represents all upstream dependencies of a release and is resolved using a dependency resolver such as \Code{pip}~\cite{pip} or \Code{Poetry}~\cite{Poetry}.

\begin{figure}
    \centering
    \includegraphics[width=0.85\linewidth]{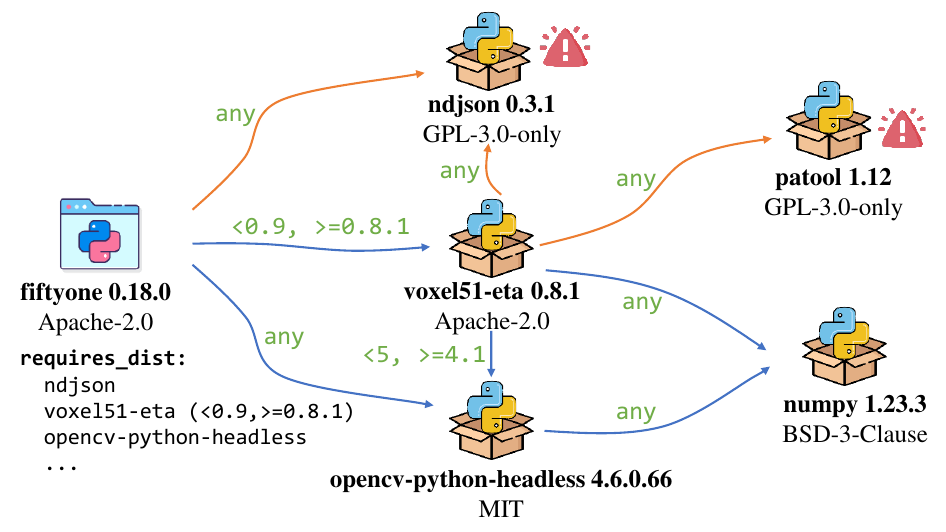}
    \vspace{-2mm}
    \caption{
    License incompatibilities in \texttt{fiftyone 0.18.0} when it is released. 
    }
    \label{fig:example}
    \vspace{-5mm}
\end{figure}

For example, Figure~\ref{fig:example} illustrates a part of the dependency graph for \Code{fiftyone 0.18.0} when it is released on November 10th, 2022. 
We can observe that \Code{fiftyone 0.18.0} depends on two GPL-3.0-licensed releases, i.e., \Code{ndjson 0.3.1} and \Code{patool 1.12}. 
However, \Code{fiftyone 0.18.0} itself is licensed under Apache 2.0, which violates the requirement of GPL 3.0 that any of its dynamically linked derivative work should be also licensed under a GPL license (as interpreted by the Free Software Foundation~\cite{GPLFAQ}).
Such license incompatibilities can happen for many reasons, including but not limited to:
1) developers may pay insufficient attention to OSS licensing or have insufficient knowledge about OSS licensing~\cite{DBLP:journals/cacm/SojerH11, DBLP:journals/ese/AlmeidaMWH19}; 
2) dependency graphs dynamically change over time~\cite{DBLP:conf/icse/LiuCF00022} and packages may change licenses in new releases~\cite{DBLP:conf/iwpc/VendomeLBPGP15, DBLP:conf/icsm/VendomeVBPGP15};  
3) developers may only manage direct dependencies, overlooking or lacking enough control over transitive dependencies~\cite{DBLP:conf/ccs/PashchenkoVM20}. 

Past research has revealed the prevalence of license incompatibilities in npm and RubyGem~\cite{DBLP:journals/jip/QiuGI21, DBLP:conf/icsr/MakariZR22} 
and techniques are proposed to detect incompatibilities~\cite{LicenseCheck,LicenseFinder,DBLP:journals/tosem/XuGFLLJ23,DBLP:conf/icse/GermanH09,DBLP:conf/ccs/DuanBXKL17,DBLP:conf/kbse/BurgDMDGH14, DBLP:journals/software/GermanP12}. 
An earlier study~\cite{DBLP:conf/icse/GermanH09} provided guidance on reusing OSS components to avoid license incompatibilities.
However, to the best of our knowledge, other packaging ecosystems are understudied and little is known about how developers remediate license incompatibilities \textit{in the dependency graph}.
Such knowledge is important for the design of tools to support this process.

To bridge the aforementioned gap, we begin with a large-scale empirical study in the PyPI ecosystem, one of the most thriving packaging ecosystems in recent years. 
To enable this study, we build an up-to-date dataset containing licensing and dependency information of 3,622,711 releases from 438,967 PyPI packages. 
Our study answers these research questions:
\vspace{-0.25mm}
\begin{itemize}
    \item \textbf{RQ1}: \textit{What is the distribution of licenses and how does licensing evolve in the PyPI ecosystem?}
    \item \textbf{RQ2:} \textit{What is the distribution of license incompatibilities in the dependency graphs of PyPI releases?} 
    \item \textbf{RQ3:} \textit{How do PyPI package developers respond to and remediate license incompatibilities in practice?} 
\end{itemize}
\vspace{-0.25mm}

Inspired by our findings, we propose \ToolName, an \underline{S}MT-solver-based \underline{i}ncompatibility remediator for \underline{l}i\underline{cen}s\underline{e}s in the dependency graph. 
Given a release and its dependency graph with one or more license incompatibilities, \ToolName~1) finds alternative licenses that are compatible with the dependency graph, and 2) searches for alternative graphs with no license incompatibilities and minimal changes compared to the original graph (i.e., indicating minimal remediation costs).
The results are aggregated as a report of recommended remediations (i.e., migrations, removals, version pinnings, or license changes) for developers to consider and choose.
Our evaluation shows that the results of \ToolName~can match the remediations proposed by developers in 19 historical real-world cases except when the migration is not covered by an existing knowledge base~\cite{guMigration}.
We further identify and report license incompatibilities that are still present in nine popular PyPI packages, five of which have been confirmed and remediated by package developers following one of the \ToolName's suggestions. 

In summary, the contributions of this paper are as follows:
\vspace{-0.25mm}
\begin{itemize}
    \item We build an up-to-date dependency and licensing dataset for the PyPI ecosystem, laying the foundation for license incompatibility analysis and remediation.
    \item We conduct the first large-scale empirical study to confirm the prevalence of license incompatibilities in PyPI
    and reveal developers' remediation practices.
    \item We design and evaluate a novel SMT-solver-based approach, \ToolName, for recommending actions to remediate license incompatibilities in Python dependency graphs. 
\end{itemize}


\section{Related Work}
\vspace{-0.25mm}

OSS licenses and licensing are studied in both software engineering and information system research. 
We review related work in three main realms: license identification, license usage and evolution, and license incompatibility detection.

\textbf{License Identification.}
The first step of any license-oriented research is the identification of licenses and/or license terms in OSS, which can be difficult in the absence of clean and curated data sources.
Therefore, researchers have proposed various approaches to identify licenses, or some specific license terms, from source code, binary files, or text~\cite{DBLP:conf/msr/Gobeille08, DBLP:journals/ase/TuunanenKK09, DBLP:conf/kbse/GermanMI10, DBLP:conf/msr/PentaGA10, DBLP:conf/kbse/LiuHGN19}.
There are also open-source tools for this purpose, such as \Code{ScanCode}~\cite{ScanCode} and \Code{Licensee}~\cite{Licensee}.
To facilitate the automated processing of OSS licensing information, the Linux Foundation proposed the Software Package Data Exchange (SPDX) standard in which a list of standard license identifiers is defined~\cite{SPDX}.

\textbf{License Usage and Evolution.}
Di Penta et al.~\cite{DBLP:conf/icse/PentaGGA10} studied the licensing evolution of six OSS systems, concluding that they underwent frequent and substantial changes with variable patterns.
Comino and Manenti~\cite{DBLP:journals/iepol/CominoM11} proposed a model to explain the commercial benefits of dual-licensed OSS.
Vendome et al.~\cite{DBLP:conf/iwpc/VendomeLBPGP15, DBLP:conf/icsm/VendomeVBPGP15} conducted a large-scale mixed-method study on 16,221 Java projects; they discovered a clear trend toward the use of less restrictive licenses mainly for facilitating reuse.
In the context of JavaScript projects, studies analyzed the use of non-approved OSI licenses~\cite{DBLP:conf/msr/Meloca0BMPWG18} and multi-licensing~\cite{DBLP:journals/ese/MoraesPWSP21}.

\textbf{License Incompatibility Detection.}
Perhaps the most important topic in OSS licensing is to check if some software is legally compliant with all the OSS it depends on, as violations can lead to legal, monetary, and ethical consequences~\cite{rosen2005open, DBLP:journals/jmis/SojerAKH14}.

Licensing issues can manifest in many ways (see a comprehensive taxonomy in Vendome et al.~\cite{DBLP:conf/icse/VendomeGPBVP18}), but most research effort is focused on checking license incompatibilities between common, known OSS licenses.
German et al.~\cite{DBLP:conf/icse/GermanH09} developed a model for license incompatibility and performed case studies on how different software systems address incompatibilities.
Further studies proposed approaches to understand and check license incompatibilities in the Fedora Linux distribution~\cite{DBLP:conf/iwpc/GermanPD10}, Android apps~\cite{DBLP:conf/ccs/DuanBXKL17}, and Java applications~\cite{DBLP:conf/kbse/BurgDMDGH14, DBLP:journals/software/GermanP12}.
Kapitsaki et al.~\cite{DBLP:journals/jss/KapitsakiKT17} proposed a general process based on SPDX.
Wolter et al.~\cite{tosem2022Inconsistencies} studied license inconsistencies within GitHub repositories, finding that many of the most popular ones do not fully declare all the licenses found in their source code.  

In the context of packaging ecosystems, Qiu et al.~\cite{DBLP:journals/jip/QiuGI21} find that 0.644\% of npm packages have license incompatibilities and developers face difficulties in managing them.
Considering more licenses and the entire ecosystem, Makari et al.~\cite{DBLP:conf/icsr/MakariZR22} find that 7.3\% of npm packages and 13.9\% of RubyGem packages contain license incompatibilities. 
Pfeiffer~\cite{DBLP:journals/corr/abs-2203-01634} studied incompatibilities caused by the AGPL license in seven ecosystems, concluding that incompatibilities are present in all ecosystems, among which PyPI and Maven packages are most risky.

Other studies explored the possibility of using fine-grained analysis on license terms to find incompatibilities in arbitrary licenses, using argumentation system~\cite{DBLP:conf/icail/Gordon11} or learning-based approaches~\cite{DBLP:journals/tosem/XuGFLLJ23}.
For example, Xu et al.~\cite{DBLP:journals/tosem/XuGFLLJ23} proposed \textsc{LiDetector}, an NLP-based method to interpret any OSS license and detect incompatibilities. 
Their analysis of 1,846 GitHub projects revealed that 72.91\% of them have license incompatibilities, but they did not consider license incompatibilities \textit{in the dependency graph}.
Researchers also studied the developers' understanding of OSS licensing~\cite{DBLP:journals/cacm/SojerH11, DBLP:journals/ese/AlmeidaMWH19, DBLP:journals/jss/PapoutsoglouKGA22}, proposed license recommendation tools~\cite{DBLP:journals/tse/KapitsakiC21, xulicenserec, ChooseALicense}, and investigated the impact of OSS licensing on different topics~\cite{DBLP:journals/isr/StewartAM06, DBLP:journals/jmis/SenSN09, DBLP:journals/jais/SojerH10, DBLP:journals/jmis/SojerAKH14}.

To the best of our knowledge, none of the previous studies have investigated how packaging ecosystem developers remediate license incompatibilities in the dependency graph of a specific package.
Such understanding is critical for the design of automated tools to address developers' demand in remediating such incompatibilities (as shown in Qiu et al.~\cite{DBLP:journals/jip/QiuGI21}).
Among different packaging ecosystems, PyPI is understudied in OSS licensing (the only study on PyPI~\cite{DBLP:journals/corr/abs-2203-01634} investigated only the AGPL license) but highly popular (currently the 3rd largest packaging ecosystem with rapid growth~\cite{ModuleCount}).
This motivates us to instantiate our study in the PyPI ecosystem.

\section{The PyPI Dependency \& Licensing Dataset}
\label{sec:dataset}

To provide a foundation for license incompatibility analysis and remediation in the dependency graph, we build a dataset with the dependency and licensing information of the entire PyPI ecosystem as of November 2022.
In this Section, we will describe the dataset and its construction process in detail.

\subsection{PyPI Dependency Data}

\subsubsection{Data Collection}

We begin with a complete PyPI distribution metadata dump obtained from the official dataset hosted on Google BigQuery~\cite{PyPIBigQuery} in November 2022.
The dump contains 438,967 packages with 3,622,711 different releases, and each release may have multiple distributions (e.g., intended for different operating systems or Python versions). 
For each distribution, the metadata provides a \Code{requires\_dist} field specifying other packages required by this distribution, optionally with version constraints and extra markers (as defined by PEP 508~\cite{PEP508}, see an example in Figure~\ref{fig:example}). 
We observe that for the same release, the \Code{requires\_dist} fields of different distributions are almost always consistent.\footnote{
Specifically, among the top 5000 most downloaded PyPI packages~\cite{PyPITop5000} (which we will also use for the empirical study, Section~\ref{sec:study-subjects}), only 0.28\% of their releases have inconsistent \texttt{requires\_dist}s in different distributions.
}
For convenience, we arbitrarily select the \Code{requires\_dist} from one distribution as the dependencies of a particular release.

\subsubsection{Dependency Resolution}
\label{sec:DepRes}

The \Code{requires\_dist} field only encodes a \textit{specification} of direct dependencies which is a list of requirement strings~\cite{core-metadata}.
Using a dependency solver (e.g., \Code{pip}~\cite{pip} or \Code{Poetry}~\cite{Poetry}), the specification can be solved into a \textit{concrete} dependency graph, with all dependencies (direct and transitive) and their versions.
Unfortunately, the relationship between dependency specifications and dependency graphs is loose: the same specification can result in different dependency graphs at different times due to new package releases, flexible version constraints, and changes in the dependency solver~\cite{DBLP:conf/icse/LiuCF00022, DBLP:conf/kbse/WangWSSL22, pipBacktracking}.
For the purpose of longitudinal analysis, we need to restore the dependency graph of a specific release at any past time of interest.
Thus, we implement a custom dependency solver following the algorithm described in Wang et al.~\cite{DBLP:conf/icse/WangW0WLWYCX020}, which imitates the breadth-first search behavior of \Code{pip} but ignores dependency conflicts and backtracking~\cite{pipBacktracking}.

To evaluate the extent to which this dependency solver can imitate \Code{pip}, we collect packages with $\ge 1$ non-optional direct dependency from the top 5000 most downloaded PyPI packages~\cite{PyPITop5000}, resulting in 825 packages. 
For each package $p$, we use our solver to solve a dependency graph $G_{ours}$ at the current time for its latest release.
Then, we run \Code{pip install} $p$ in a clean virtual environment to get a ground truth dependency graph $G_{pip}$ solved by \Code{pip} and compute precision \& recall as:
\begin{align*}
    Precision(p) &= |G_{ours} \cap G_{pip}|\ /\ |G_{ours}|\\
    Recall(p) &= |G_{ours} \cap G_{pip}|\ /\ |G_{pip}|
\end{align*}
Among the 825 packages, we obtain an average precision of 0.9715 and an average recall of 0.9390, indicating a very high degree of match between the results of the two solvers.
The mismatches can happen for various reasons, such as the four-month lag between our dump and the experiment time, \Code{pip}'s backtracking behaviors~\cite{pipBacktracking}, etc.
Still, our custom dependency solver is orders of magnitude faster than \Code{pip} because it directly queries our metadata dump (instead of interacting with PyPI APIs and downloading a lot of release files).
It also supports resolving dependency graphs at any historical time of interest, which is not possible using \Code{pip}.
Using this solver, we compute a historical dependency graph for each release \textit{at its upload time}, which we will use for our empirical study.\footnote{
Note that this approach ignores development dependencies and optional dependencies, computing only the dependency graph that is always distributed with the package (e.g., when a \texttt{pip install} is executed).
This means that any license incompatibilities, especially those related to redistribution, would be highly problematic if present in this dependency graph. 
}

\subsection{PyPI Licensing Data}
\label{sec:pypi-licensing}

\subsubsection{Data Collection}
\label{sec:licensing-data-collection}

The licensing information of a release can be found in three possible data sources:
\begin{itemize}
    \item The \Code{license} field in its distribution metadata. It has two notable limitations: 1) its value is left to the discretion of individual developers without a uniform format (e.g., an Apache 2.0 licensed package can have values like \Code{"Apache v2"}, \Code{"Apache Version 2"}, \Code{"Apache 2"}, or even the complete license text; 2) 31.9\% of the releases do not have this field in its distribution metadata. 
    \item The \Code{classifier} field in the metadata may contain pre-defined license identifiers that can be easily mapped into SPDX identifiers~\cite{SPDX}. This data source is validated by PyPI and can serve as ground truth, but even fewer (13.8\%) releases have license tag(s) in \Code{classifier}. 
    \item The wheel distribution files, which can include \Code{LICENSE} and \Code{README} files with licensing information. However, downloading all distribution files from PyPI would  require excessive computation and network resources.
\end{itemize}

To address the limitations of these data sources, we design a multi-step cross-validation approach to get cleaned licensing information (as SPDX license identifiers) for as many releases as possible in our dataset.
For this purpose, we build a mapping between \Code{license} fields and SPDX license identifiers using all package versions with available classifier tags.
Using this mapping, we build another mapping between SPDX license identifiers and common keywords in the \Code{license} fields, including name keywords, version keywords, ``must-not-have'' keywords, and ``must-have'' keywords.
The two mappings are intended to ``cross-validate'' \Code{license} fields using the ground truth available from the license classifier tags.

For each release, if it already contains a license identifier in the \Code{classifier} field, we just convert it to the SPDX identifier.
Then, for each of the remaining releases with a \Code{license} field, we retrieve the most frequent SPDX license identifier corresponding to the value of this field using the first mapping.
If the above retrieval does not work, we use the second keyword mapping (which is looser) to map the \Code{license} field into one of the SPDX license identifiers.
If the previous steps fail, or if the release does not have a \Code{license} field, we download its distribution file and scan the \Code{LICENSE} and \Code{README} files using \Code{ScanCode}~\cite{ScanCode}, a widely used license detection tool. 
Finally, if all attempts fail to resolve into an SPDX identifier, we mark the license as \Code{Unrecognizable}.

By applying the above approach to the 3,622,711 releases in our dataset, we get licensing information from classifier tags, the \Code{license} field, and distribution files for 500,457 (13.8\%), 2,465,863 (68.1\%), 135,590 (3.7\%) releases respectively, leaving 520,801 (14.4\%) releases with \Code{Unrecognizable} licensing.

To evaluate the effectiveness of this license identification approach,
we randomly sample 385 releases from the population of 3,622,711 releases (95\% confidence level, 5\% confidence interval~\cite{SampleSizeCalculator}).
Then, we manually check whether the licenses identified by our approach can match different sources of information, including 1) GitHub repositories), 2) \Code{LICENSE} files in the distribution, and 3) the \Code{license} field.
Among the 385 samples, our approach returns \Code{Unrecognizable} for 51 of them (13.2\%). 
Among the remaining 334 samples, 323 match other sources of information, resulting in an accuracy of 96.7\% (323 / 334). 
For the 11 misidentified samples, six are due to users providing incorrect licensing information in the metadata, four are because users omit the versions of their license in the \Code{license} field, and one is due to dual licensing.
Among the 51 samples with \Code{Unrecognizable} licensing, ten have been removed from PyPI at the time of inspections, 
39 do not have licensing information in all sources, and five are early releases of a package (developers may only consider licensing until official release~\cite{DBLP:conf/icsm/VendomeVBPGP15}). 
Two samples have custom licenses that are not covered by existing license identifiers. 
Finally, for one sample, there is no sufficient information in both the \Code{license} field and the \Code{LICENSE} file to determine the specific license for the release.
To summarize, the evaluation results demonstrate that our approach is able to identify licensing information in the majority of cases except when the data sources are noisy or dual licensing is used, but both cases are rare.
We believe that the resulting licensing information can provide a sound foundation for subsequent analyses.

\subsubsection{Finding License Incompatibilities}

Inspired by previous works~\cite{DBLP:journals/jss/KapitsakiKT17, xulicenserec}, we consider the \textit{one-way combinative incompatibility} between licenses in this paper, defined as:

\begin{defn}
 (\textit{License Incompatibility}) License $A$ is one-way incompatible with license $B$ if and only if it is infeasible to distribute derivative works of $A$-licensed software under $B$. 
\end{defn}



For example, GPL 3.0 is one-way incompatible with Apache 2.0  because the derivative works of GPL-3.0-licensed software 
cannot be distributed under Apache 2.0 (but the reverse is feasible, and thus \textit{one-way} incompatible). 
On the other hand, Apache 2.0 and GPL 2.0 are incompatible in both ways because they have conflicting terms about patents~\cite{ApacheV2andGPLv2}.

This definition fits well in the context of packaging ecosystems because a package can be considered the derivative work of its dependencies (according to the Free Software Foundation (FSF) but there are some controversies~\cite{WhatIsDeriveWork, ProsDynaLinkDeriveWork, ConsDynaLinkDeriveWork}). 

We compute all one-way incompatible license pairs using the license compatibility matrix proposed by Xu et al.~\cite{xulicenserec}, in which they analyzed the compatibility between licenses along 19 dimensions of terms such as copyleft, trademark grants, and patent grants. 
We choose this matrix for three reasons. 
First, it is the largest available license compatibility data to the best of our knowledge, compromising compatibility relationships between 63 licenses. 
Second, to ensure popularity and representativeness, all the licenses are: 1) certified by FSF or OSI~\cite{OSIApproved}; 2) not obsolete
(e.g., Apache-1.1); 
3) not restricted to specific domains,
software, or authors (e.g., IPA is a font license). 
Third, the 63 licenses can cover 99.4\% of releases of which the license information has been obtained in our dataset.

Using these incompatible license pairs, we identify incompatible dependencies for each release based on the dependency graphs computed in Section~\ref{sec:DepRes}.

\subsection{Dataset Overview}
\label{sec:def}

To summarize, our dataset contains 438,967 PyPI packages and 3,622,711 releases from the entire PyPI ecosystem as of November 2022.
For each release, the dataset offers 1) an SPDX license identifier, 2) a list of direct dependencies and their version constraints, 3) a dependency graph at its upload time, and 4) a list of incompatible dependencies.
The dataset is stored as a MongoDB collection occupying 3.45GB of storage space with built-in compression.
As most of the dataset construction process is automated (except building the keyword mapping in Section~\ref{sec:licensing-data-collection}), the dataset can be easily updated using the latest PyPI BigQuery dataset.
To the best of our knowledge, this is the \textit{first} dataset of dependency and licensing information in the entire PyPI ecosystem.
We will discuss the limitations of this dataset in Section~\ref{sec:limitation}.

\section{Empirical Study}
\label{sec:long-study}  

\subsection{Research Questions}

The goal of this empirical study is to provide evidence about license incompatibilities and their remediation practices in the PyPI ecosystem.
Such evidence can help the design of automated tools supporting remediation in dependency graphs.
Toward this goal, we ask the following research questions:

\begin{itemize}
    \item \textbf{RQ1:} \textit{What is the distribution of licenses and how does licensing evolve in the PyPI ecosystem?}
    
    \textbf{Rationale.} 
    This \textbf{RQ} aims to provide an overview of licenses and licensing evolution in the PyPI ecosystem.
    We are especially interested in the prevalence and evolution of restrictive licenses as they are most likely to introduce license incompatibilities. 
    Although the same question has been answered in other contexts~\cite{DBLP:conf/iwpc/VendomeLBPGP15, DBLP:conf/icsr/MakariZR22}, it has not been answered in PyPI yet, motivating us to ask this \textbf{RQ}.
    \item \textbf{RQ2:} \textit{What is the distribution of license incompatibilities in the dependency graphs of PyPI releases?}
    
    \textbf{Rationale.} 
    Due to the prevalence of license incompatibilities in npm and RubyGem~\cite{DBLP:journals/jip/QiuGI21, DBLP:conf/icsr/MakariZR22}, this \textbf{RQ} intends to confirm, in PyPI, the prevalence of license incompatibilities.
    We are also interested in their positions in the dependency graph (direct or transitive), and their degree of connectivity with other nodes in the dependency graph, which may indicate possible difficulties in remediation. 
    \item \textbf{RQ3:} \textit{How do PyPI package developers respond to and remediate license incompatibilities in practice?}
    
    \textbf{Rationale.}
    The goal of this \textbf{RQ} is to uncover the challenges that developers face when attempting to remediate license incompatibilities and to explore common remediation strategies discussed by developers. 
    Such understanding is vital for the design of supportive tools, especially in the design of potential solution spaces.
\end{itemize}

\subsection{Study Subjects}
\label{sec:study-subjects}

For \textbf{RQ1} \& \textbf{RQ2}, we consider two groups of PyPI packages:
\begin{itemize}
    \item \textsc{Top}: The top 5000 most downloaded PyPI packages~\cite{PyPITop5000}.
    This group represents widely-used Python packages for which license incompatibilities can have a huge impact;
    \item \textsc{All}: All the 438,967 PyPI packages in our dataset.
\end{itemize}
We expect a comparison to reveal the differences between popular packages and the global population in terms of their license preferences and licensing practices.
To avoid bias from packages with a large number of releases, we only select the latest release of each package in each year for all subsequent analyses (except for within-package evolution in \textbf{RQ1}).

For \textbf{RQ3}, we only focus on the \textsc{Top} group as they are more likely to have mature development practices and transparent development activities (e.g., extensively using issue trackers), without which the answering of \textbf{RQ3} would be impossible.

\subsection{Methods and Results}

\subsubsection{RQ1: License Distribution \& Evolution}

Following prior work~\cite{ DBLP:journals/jss/KapitsakiKT17, DBLP:journals/tse/KapitsakiC21, tosem2022Inconsistencies}, we classify licenses into four different categories ordered by their level of permissiveness: 
\begin{itemize}
    \item \textbf{Permissive}: Software that changes or uses existing software can be licensed under a different license (e.g., MIT); 
    \item \textbf{Weak Copyleft}: Software that changes existing software must be licensed under the same license, but software that uses existing software (e.g., by calling APIs) does not have to (e.g., LGPL 3.0). 
    \item \textbf{Strong Copyleft}: Software that changes or uses existing software must be licensed under the same license unless an exception is specified (e.g., GPL 3.0 and AGPL 3.0); 
    \item \textbf{Unknown}: The license is \Code{Unrecognized} (Section~\ref{sec:pypi-licensing}).
\end{itemize}

Overall, widely-used PyPI packages tend to be permissive: in the \textsc{Top} group, 85.82\% have a permissive license, 4.07\% have a weak copyleft one, and 3.72\% have a strong copyleft one, leaving 6.39\% as unknown.
However, the global population is more restrictive and less recognizable: in the \textsc{All} group, the ratio of packages with a permissive, weak left, strong copyleft license is 62.14\%, 2.80\%, and 14.67\% respectively, leaving a large proportion of 20.39\% as unknown.

\begin{figure}
    \centering
    \subfigure[\textsc{Top}]{\includegraphics[width=0.48\linewidth]{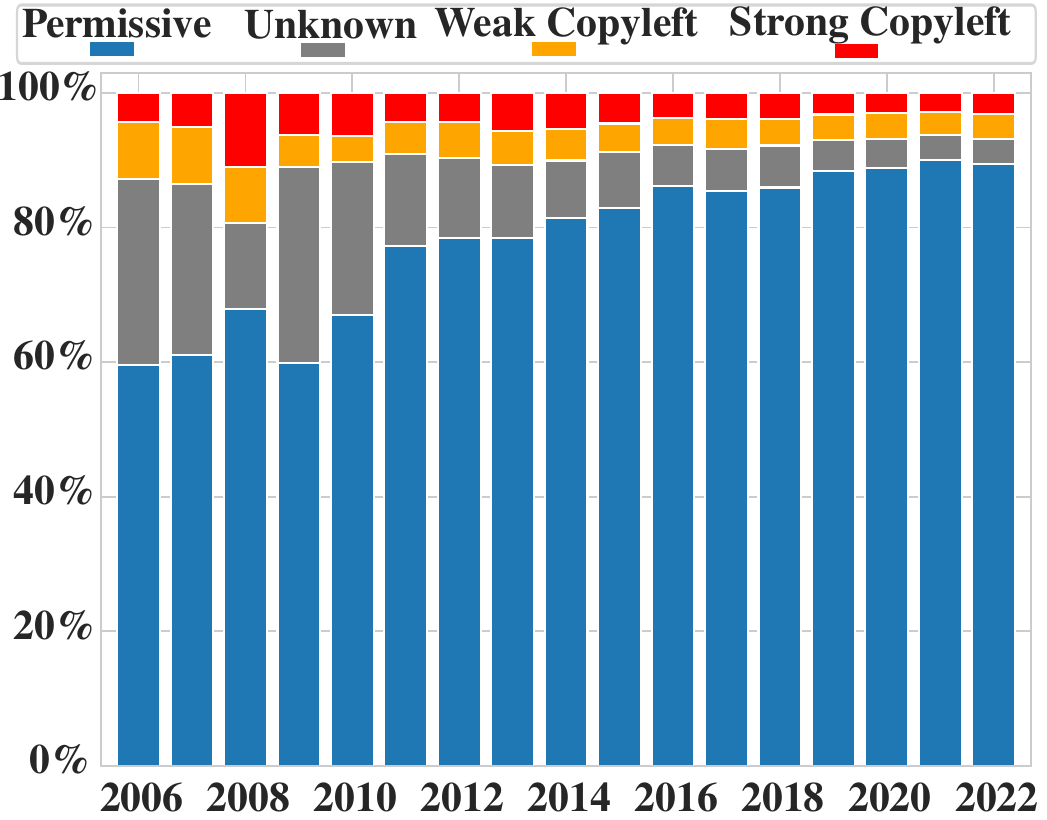}\label{fig:license_type_yearly_freq_percentage}}
    \subfigure[\textsc{All}]{\includegraphics[width=0.48\linewidth]{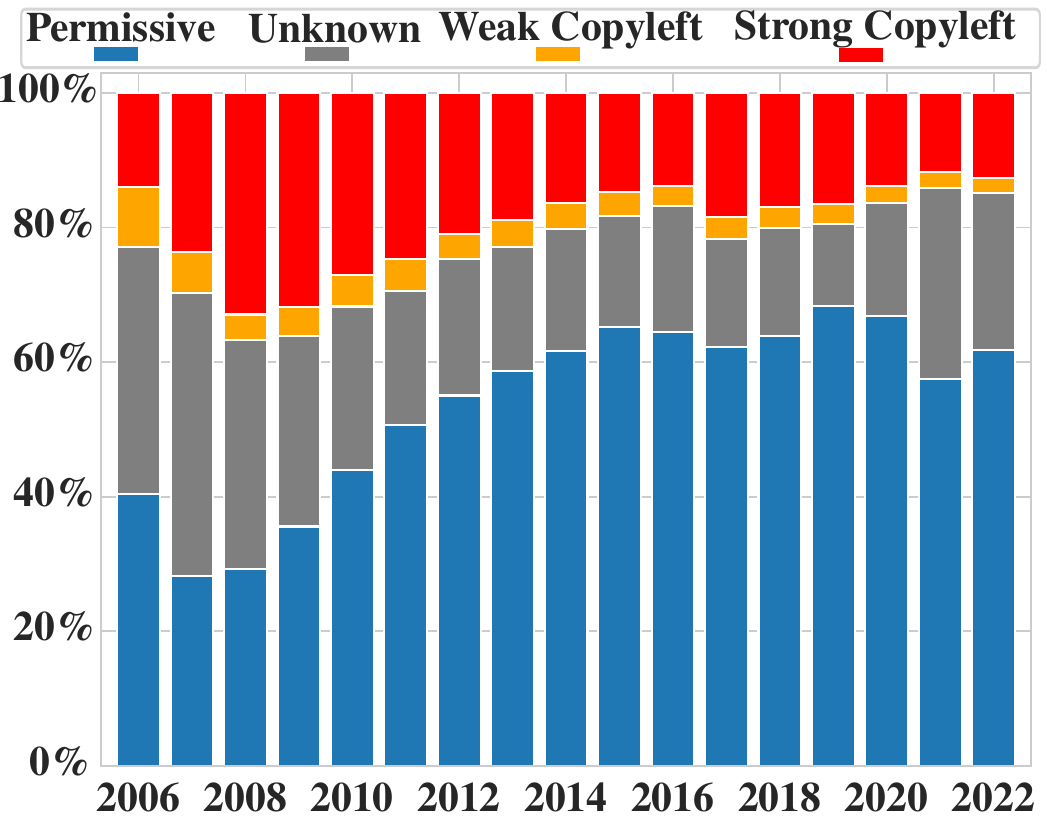}\label{fig:license_type_yearly_freq_percentage_pypi}}
    \vspace{-2mm}
    \caption{The yearly distribution of licensing categories in the two groups.}
    \label{type}
    \vspace{-4mm}
\end{figure}

We plot the yearly distribution of licensing categories in Figure~\ref{fig:license_type_yearly_freq_percentage} and \ref{fig:license_type_yearly_freq_percentage_pypi}.
We can observe that permissive licenses are not only the most common but also increasingly popular over the years in both groups.
However, as of 2022, packages with strong copyleft licenses in \textsc{All} still constitute a significant portion (12.63\%) and 4.0x higher than that of \textsc{Top} (3.17\%). 
What's more, the proportion of the unknown category in \textsc{Top} is lower than that in \textsc{All} and is decreasing over the years.
This indicates that widely-used packages have devoted efforts to providing accurate and complete licensing information but less popular ones have not done so.

Similar to Vendome et al.~\cite{DBLP:conf/iwpc/VendomeLBPGP15}, we investigate how licensing evolves \textit{within packages}. 
We confirm that licensing changes are not uncommon in PyPI packages (just as other OSS~\cite{DBLP:conf/icse/PentaGGA10, DBLP:conf/iwpc/VendomeLBPGP15}): 
in the \textsc{Top} group, 425 (9.10\%) packages have undergone one licensing change, and 87 (1.86\%) packages have undergone two or more changes. 
This is significantly higher than that in \textsc{All} (3.04\%). 
Most licensing changes are between licenses in the same level of permissiveness (63.74\% in \textsc{Top} and 56.20\% in \textsc{All}).
In \textsc{Top}, there is a tendency toward using more permissive licenses (27.66\%) but changing toward less permissive ones is less frequent (8.60\%).
In \textsc{All}, licensing changes in both directions are common (26.15\% toward more permissive and 17.65\% toward less permissive).

\begin{small}
\begin{summary-rq}
\textbf{Answers for RQ1:}
In the PyPI ecosystem, 85.82\% of the \textsc{Top} packages have a permissive license, but strong copyleft licenses are also present (3.72\% among \textsc{Top} and 14.67\% among \textsc{All}).
10.96\% of the \textsc{Top} packages and 3.04\% of \textsc{All} have undergone at least one licensing change.
Although many licensing changes are within the same level of permissiveness, a non-negligible portion is toward more restrictive ones (17.65\% among \textsc{All}).

\vspace{-4pt}
\noindent\rule{\textwidth}{0.8pt}

\textbf{Implications:}
The risk of license incompatibilities could be high in PyPI due to the presence of strong copyleft licenses.
We also confirm that licensing changes are common in PyPI packages, among which changing toward more restrictive licenses could be especially problematic for the downstream packages.
To take licensing changes into consideration, a precise and versioned dependency graph is necessary for license incompatibility analysis.
\end{summary-rq}
\end{small}

\subsubsection{RQ2: License Incompatibility Distribution}


In Section~\ref{sec:def}, we have resolved a dependency graph for each release at its upload time and checked whether the licenses of all its dependencies in the graph are compatible with the license of this release. 
If any incompatibility is detected, we label the release as \Code{Incompatible}; if all dependencies have compatible licenses, we label the release as \Code{Compatible}; otherwise, (i.e., there is at least one dependency with \Code{Unrecognizable} license), we label the release as \Code{Unknown}.


\begin{table}
\footnotesize
\centering
\caption{The license compatibility status of PyPI releases}
\vspace{-0.2cm}
\label{tab:incom}
\begin{tabular}{lrrrr}
\toprule
&\multicolumn{2}{c}{\textsc{Top} (10,282 releases)}& \multicolumn{2}{c}{\textsc{All} (271,811 releases)}\\
Compatibility Label & Count& Percentage& Count& Percentage\\
\midrule
\texttt{Compatible} &	5,731 &	55.64\% &	114,135 &	41.99\%\\
\texttt{Incompatible} &	202&	1.96\% &	19,772 &	7.27\%\\
\texttt{Unknown} &	4,349&	42.30\%&	137,904 &	50.74\%\\
\bottomrule
\end{tabular}
\vspace{-0.3cm}
\end{table}

\begin{table}
\footnotesize
\centering
\setlength{\tabcolsep}{1.25mm}
\caption{The cumulative distribution of dependency graph metrics for  all incompatible dependencies}
\vspace{-0.2cm}
\label{tab:metric}
\begin{tabular}{lccccccc}
\toprule
& & $= 0$ & $\le 1$ & $\le 2$ & $\le 3$ & $\le 4$ & $\le 5$\\
\midrule
\multirow{2}{*}{\texttt{Depth}} & \textsc{Top} & - & 74.0\% & 96.2\% & 100\% & - &-\\
& \textsc{All} & - & 38.7\% & 71.8\%& 89.1\% & 95.2\% & 97.6\% \\
\midrule
\multirow{2}{*}{\texttt{In-degree}} & \textsc{Top} & - &95.9\% & 100\%& - & - & -\\
& \textsc{All} &-&75.4\% & 87.7\% & 91.9\% & 94.2\% & 95.8\%\\
\midrule
\multirow{2}{*}{\texttt{Out-degree}} & \textsc{Top} &60.8\% &68.7\% & 81.9\% & 90.2\% & 97.4\% & 97.7\% \\
& \textsc{All} &45.6\%& 57.9\% & 63.7\% & 68.6\%& 74.1\% & 79.0\% \\
\bottomrule
\end{tabular}
\vspace{-0.3cm}
\end{table}

As we study license incompatibilities introduced by dependencies, we exclude releases without dependencies, leaving 10,282 releases in the \textsc{Top} group (3,068 packages) and 271,811 releases in the \textsc{All} group (176,955 packages).
We summarize their license compatibility status in Table~\ref{tab:incom}.
We can observe that license incompatibilities are less common among \textsc{Top}, with only 202 (1.96\%) of the releases being \Code{Incompatible} (92 packages). 
However, this proportion is significantly higher among \textsc{All}, with 19,772 (7.27\%) being \Code{Incompatible}. 
This indicates that license incompatibilities are not uncommon in PyPI ecosystem and much more common (3.7x) in less popular packages than widely-used packages.

In the dependency graph of a release, license incompatibility can be caused by both direct dependencies and transitive dependencies. 
The latter is more difficult to remediate because:
1) transitive dependencies are required by other dependencies and developers have limited control over them;
2) their remediation can trigger a ripple effect due to edges in the graph.
Therefore, to gain a better understanding of this problem, we are interested in the \textit{location} of license incompatibilities in dependency graphs.
For each license incompatibility, we compute the following metrics in the dependency graph:
\begin{itemize}
    \item \textbf{Depth:} The shortest distance between the incompatible dependency and the root node. Direct dependencies have a depth of one. A high depth means a long dependency chain needs to be addressed during remediation. 
    \item \textbf{In-degree:} The number of packages in the dependency graph directly depending on the incompatible dependency, which needs to meet the version constraints for all of them. In-degree characterizes the number of constraints that need to be considered during remediation.
    \item \textbf{Out-degree:} The number of packages that the incompatible dependency directly depends on. Out-degree characterizes the number of dependencies that could be impacted when remediating the compatibility issue. 
\end{itemize}

Table~\ref{tab:metric} shows the cumulative distribution of these metrics for incompatible dependencies in the dependency graph. 
In total, there are 265 and 46,237 incompatible dependencies in the \textsc{Top} and \textsc{All} groups, respectively (a release may have multiple incompatible dependencies).
We find that incompatible dependencies are more likely to be in a complex position among \textsc{All} compared with \textsc{Top}. 
Among \textsc{Top}, 26.0\% of them come from transitive dependencies (i.e., depth $\ge$2) while the percentage rises to 61.3\% among \textsc{All}.
5,032 (10.9\%) of them in \textsc{All} have a depth of at least four in the dependency graph, and 5,681 (12.3\%) of them have an in-degree greater than or equal to three.
However, among \textsc{Top}, all cases of license incompatibilities caused by transitive dependencies are limited to the second or third layer of the dependency graph, with an in-degree of either one or two.
Moreover, among \textsc{All}, the mean of the out-degree for incompatible dependencies in the dependency graph is 3.93, whereas in \textsc{Top}, it is only 1.06. 

In other words, license incompatibilities are sophisticated for many releases in the PyPI ecosystem.
They may be caused by incompatible transitive dependencies, some of which are deeply nested with many dependencies and dependents. 
This means that the remediation of these incompatibilities requires addressing many other interrelated dependencies, necessitating a method that can identify feasible solutions from a global perspective considering the entire dependency graph.

\begin{small}
\begin{summary-rq}
\textbf{Answers for RQ2:}
In the entire PyPI ecosystem, a significant proportion of releases (7.27\%) have license incompatibilities.
Although most incompatible dependencies (74.0\%) are direct dependencies in dependency graphs of \textsc{Top} packages, 61.3\% of them in that of \textsc{All} are transitive dependencies that may reside in deep and sophisticated dependency graph positions.

\vspace{-4pt}
\noindent\rule{\textwidth}{0.8pt}

\textbf{Implications:}
License incompatibilities form a significant problem in the PyPI ecosystem.
Remediating license incompatibilities in transitive dependencies requires searching for a feasible solution from a global perspective in the entire dependency graph.
\end{summary-rq}
\end{small}

\subsubsection{RQ3: License Incompatibility Remediation in Practice}

\begin{table*}[]
    \tabcolsep=0.16cm
    \centering
    \footnotesize
    \caption{License Incompatibilities and Their Remediations. $\bigstar$ Marks the Final Remediation Taken By Developers.}
    \vspace{-2mm}
    \begin{tabular}{cllllll}
    \toprule
        & Package & License & Incompatible Dependency & License & Issue(s) \& PR(s) & Proposed Remediation(s)  \\
    \midrule
        \parbox[t]{-3mm}{\multirow{21}{*}{\rotatebox[origin=c]{90}{\textbf{Identified in RQ3}}}} & \texttt{ansible-lint} & MIT & \texttt{ansible} & GPL 3.0 & \href{https://github.com/ansible/ansible-lint/issues/1188}{\#1188}, \href{https://github.com/ansible/ansible-lint/pull/1882}{\#1882} &  $\bigstar$Change Own License\\
        & \texttt{apache-airflow} & Apache 2.0 & \texttt{mysql-connector-python} & GPL 3.0 & \href{https://github.com/apache/airflow/issues/9898}{\#9898}, \href{https://github.com/apache/airflow/issues/10667}{\#10667} & Migration, $\bigstar$No Remediation\\
        & \texttt{cvxpy} & Apache 2.0 & \texttt{ecos} & GPL 3.0 &\href{https://github.com/cvxpy/cvxpy/issues/313}{\#313} & Migration, $\bigstar$No Remediation \\
        & \texttt{dvc} & Apache 2.0 & \texttt{grandalf} & GPL 2.0 & \href{https://github.com/iterative/dvc/issues/1115}{\#1115} & $\bigstar$No Remediation \\
        & \texttt{fbprophet} & 3-Clause BSD & \texttt{lunardate} & GPL 3.0 &  \href{https://github.com/facebook/prophet/issues/1069}{\#1069}, \href{https://github.com/facebook/prophet/pull/1091}{\#1091} & $\bigstar$Migration, Removal\\
        & \texttt{fbprophet} & 3-Clause BSD & \texttt{pystan} & GPL 3.0 & \href{https://github.com/facebook/prophet/issues/1045}{\#1045}, \href{https://github.com/facebook/prophet/issues/1221}{\#1221}  & $\bigstar$Migration \\
        & \texttt{fiftyone} & Apache 2.0 & \texttt{ndjson} & GPL 3.0 & \href{https://github.com/voxel51/fiftyone/pull/2864}{\#2864}, \href{https://github.com/voxel51/eta/pull/590}{\texttt{eta}\#590} & $\bigstar$Migration\\ 
        & \texttt{fiftyone} & Apache 2.0 & \texttt{patool} & GPL 3.0 & \href{https://github.com/voxel51/fiftyone/pull/2864}{\#2864}, \href{https://github.com/voxel51/eta/pull/590}{\texttt{eta}\#590} & $\bigstar$Migration\\
        & \texttt{halo} & MIT & \texttt{cursor} & GPL 3.0 & \href{https://github.com/manrajgrover/halo/issues/118}{\#118}, \href{https://github.com/manrajgrover/halo/pull/147}{\#147} & Pin Version, Migration, $\bigstar$Removal \\
        & \texttt{jiwer} & Apache 2.0 & \texttt{levenshtein} & GPL 3.0 & \href{https://github.com/jitsi/jiwer/issues/69}{\#69}, \href{https://github.com/jitsi/jiwer/pull/71}{\#71} & $\bigstar$Migration\\
        & \texttt{mitmproxy} & MIT & \texttt{html2text} & GPL 3.0 & \href{https://github.com/mitmproxy/mitmproxy/issues/2572}{\#2572}, \href{https://github.com/mitmproxy/mitmproxy/pull/2573}{\#2573} & $\bigstar$Removal\\
        & \texttt{netcdf4} & MIT & \texttt{cftime} & GPL 3.0 & \href{https://github.com/Unidata/netcdf4-python/issues/1000}{\#1000}, \href{https://github.com/Unidata/netcdf4-python/issues/1073}{\#1073} & $\bigstar$Negotiation, Pin Version\\
        & \texttt{orbit-ml} & Apache 2.0 & \texttt{pystan} & GPL 3.0 & \href{https://github.com/uber/orbit/issues/435}{\#435} & Migration\\
        & \texttt{pulp} & 3-Clause BSD & \texttt{amply} & EPL 1.0 & \href{https://github.com/coin-or/pulp/issues/394}{\#394} & $\bigstar$Negotiation, Removal\\
        & \texttt{pytest-pylint} & MIT & \texttt{pylint} & GPL 2.0+ & \href{https://github.com/carsongee/pytest-pylint/issues/178}{\#178} & No Remediation\\
        & \texttt{textacy} & Apache 2.0 & \texttt{fuzzywuzzy} & GPL 2.0 & \href{https://github.com/chartbeat-labs/textacy/issues/62}{\#62}, \href{https://github.com/chartbeat-labs/textacy/issues/63}{\#63} & $\bigstar$Removal\\
        & \texttt{textacy} & Apache 2.0 & \texttt{unidecode} & GPL 2.0+ & \href{https://github.com/chartbeat-labs/textacy/issues/203}{\#203} & Migration, $\bigstar$Removal\\
        & \texttt{textacy} & Apache 2.0 & \texttt{python-levenshtein} & GPL 3.0 & \href{https://github.com/chartbeat-labs/textacy/issues/203}{\#203} & Migration\\
        & \texttt{wemake-python} & MIT & \texttt{flake8-isort} & GPL 2.0 & \href{https://github.com/wemake-services/wemake-python-styleguide/issues/2481}{\#2481} & Negotiation, $\bigstar$Migration\\
        & \texttt{workalendar} & MIT & \texttt{lunardate} & GPL 3.0 & \href{https://github.com/workalendar/workalendar/issues/346}{\#346}, \href{https://github.com/workalendar/workalendar/issues/536}{\#536}, \href{https://github.com/workalendar/workalendar/pull/709}{\#709} & Change Own License, $\bigstar$Migration, Removal\\
        & \texttt{yt-dlp} & Unlicense & \texttt{mutagen} & GPL 2.0+ & \href{https://github.com/yt-dlp/yt-dlp/issues/348}{\#348}, \href{https://github.com/yt-dlp/yt-dlp/issues/2345}{\#2345} & Change Own License, Removal\\
        \midrule
        \parbox[t]{-3mm}{\multirow{9}{*}{\rotatebox[origin=c]{90}{\textbf{Reported by \ToolName}}}} & \texttt{amundsen} & Apache 2.0 & \texttt{unidecode} & GPL 2.0+ & \href{https://github.com/amundsen-io/amundsen/issues/2148}{\#2148}, \href{https://github.com/amundsen-io/amundsen/pull/2168}{\#2168} & Chg. Own Lic., $\bigstar$Migration, Removal, Pin Ver.\\
        & \texttt{cibuildwheel} & 2-Clause BSD & \texttt{bashlex} & GPL 3.0 & \href{https://github.com/pypa/cibuildwheel/issues/1484}{\#1484} & Change Own License, Removal\\
        & \texttt{glean-parser} & MPL 2.0 & \texttt{yamllint} & GPL 3.0 & \href{https://bugzilla.mozilla.org/show_bug.cgi?id=1830049}{\#1830049}, \href{https://github.com/mozilla/glean_parser/pull/578}{\#578} & Change Own License, $\bigstar$Removal\\
        & \texttt{metaflow} & Apache 2.0 & \texttt{pylint} & GPL 2.0+ & \href{https://github.com/Netflix/metaflow/issues/1377}{\#1377}, \href{https://github.com/Netflix/metaflow/pull/1378}{\#1378} & Change Own License, Migration, $\bigstar$Removal\\
        & \texttt{music-assistant} & Apache 2.0 & \texttt{unidecode} & GPL 2.0+ & \href{https://github.com/music-assistant/hass-music-assistant/issues/1220}{\#1220} & Change Own License, Migration, Removal\\
        & \texttt{optbinning} & Apache 2.0 & \texttt{ecos} & GPL 3.0+ & \href{https://github.com/guillermo-navas-palencia/optbinning/issues/242}{\#242} &  Change Own License, Removal\\
        & \texttt{pylint-gitlab} & MIT & \texttt{pylint} & GPL 2.0+ & \href{https://gitlab.com/smueller18/pylint-gitlab/-/merge_requests/15}{\#15}, \href{https://gitlab.com/smueller18/pylint-gitlab/-/issues/20}{\#20} & $\bigstar$Change Own License, Migration, Removal \\
        & \texttt{sphinx-autoapi} & MIT & \texttt{unidecode} & GPL 2.0+ & \href{https://github.com/readthedocs/sphinx-autoapi/issues/382}{\#382}, \href{https://github.com/readthedocs/sphinx-autoapi/commit/0a557fc95eb1130efb9459d403dfbc30c003024c}{0a557fc} & Chg. Own Lic., $\bigstar$Migration, Removal, Pin Ver.\\
        & \texttt{zha-quirks} & Apache 2.0 & \texttt{zigpy} & GPL 3.0 & \href{https://github.com/zigpy/zha-device-handlers/issues/2356}{\#3256} &  Change Own License, Removal\\
    \bottomrule
    \end{tabular}
    \vspace{-3mm}
    \label{tab:remediation}
\end{table*}

To answer \textbf{RQ3}, we analyze the GitHub issue trackers of the 92 packages with license incompatibilities from \textsc{Top}.
For each package, we manually find their GitHub repository and search the issue tracker using three different keywords: 1) \Code{license}; 2) the name of incompatible license (e.g., \Code{GPL}); 3) the name of the incompatible package (e.g., \Code{unidecode}).
Then, we manually identify relevant issues, pull requests (PRs), and discussions from the search results, resulting in 25 issues and eight PRs from 17 repositories.
For each repository, we find the developers' discussions and categorize the remediations (or proposed remediations) using an open-coding procedure~\cite{khandkar2009open}.
To ensure reliability and avoid bias, two authors of this paper, both with over five years of software development experience, independently performed the above steps; they later discussed and merged the results into a consensus.

The upper half of Table~\ref{tab:remediation} summarizes the 21 license incompatibilities we found and the remediations proposed or taken by developers. 
We have two immediate observations:

\textit{a) License incompatibilities happen because OSS developers lack knowledge or pay little attention to OSS licensing.} 
For example, a developer commented: 
\textit{I don't get into licensing much and hence MIT everything, thus don't know the implications of this. I will investigate this and get back.} (\href{https://github.com/manrajgrover/halo/issues/118}{\Code{halo}\#118}).

\textit{b) License incompatibilities frequently cause confusion and controversies, even among experienced OSS developers, after they are raised in an issue.}
Many issues in Table~\ref{tab:remediation} triggered lengthy discussions about whether the incompatibility really exists and whether it really matters for their projects (e.g., \href{https://github.com/coin-or/pulp/issues/394}{\Code{pulp}\#394}).
For example, a common argument is that having a GPL-3.0-licensed dependency does not result in the package becoming a ``derivative work'' of that dependency.
However, this contradicts the interpretation of FSF~\cite{ProsDynaLinkDeriveWork} and is disagreed by many other developers.
The situation is more controversial and sophisticated in some cases, such as with the presence of optional dependencies (e.g., \href{https://github.com/apache/airflow/issues/9898}{\Code{apache-airflow}\#9898}).

In 17 of the 21 cases, developers acknowledged the relevance of license incompatibilities and the necessity of remediation.
However, it can be non-trivial to find an appropriate remediation method and developers often need to evaluate multiple possibilities (as can be observed in Table~\ref{tab:remediation}).
Specifically, they considered the following remediation methods:

\textit{a) Migration (13 Incompatibilities):}
The most common remediation is to migrate the incompatible dependency to an alternative package with similar functionalities.
For example, \Code{lunardate} can be replaced with \Code{LunarCalendar} and \Code{unidecode} can be replaced with \Code{text-unidecode}.
This observation echoes prior research showing that developers migrate packages due to licensing issues~\cite{DBLP:conf/wcre/HeXMXLZ21, DBLP:conf/sigsoft/HeHGZ21}.

\textit{b) Removal (8 Incompatibilities):}
If the incompatible dependency is not used extensively, developers choose to remove the dependency and replace it with their own implementations of the desired functionality.
For example, the developers of \Code{halo} eventually decided, after lengthy discussions, to remove \Code{cursor} and re-implement based on a Stack Overflow snippet.

\textit{c) Change Own License (3 Incompatibilities):}
Some developers proposed changing their package's own license to comply with the licensing requirement of its dependency.
This remediation was finally taken by \Code{ansible-lint} as it is closely integrated with its GPL-licensed dependency, \Code{ansible}.

\textit{d) Negotiation (3 Incompatibilities):}
Another feasible option is to ask upstream developers (i.e., developers of the incompatible dependency) to change the licenses of their packages toward more permissive ones.
For example, \Code{cftime} decided to remove GPL-related code and relicense itself under MIT after a request from \Code{netcdf4} developers 
 (\href{https://github.com/Unidata/cftime/issues/116}{\Code{cftime}\#116}).

\textit{e) Pin Version (2 Incompatibilities):}
In the case of \Code{cursor} and \Code{cftime}, the two packages were initially released under a permissive license but changed their license in a new release.
To remediate this, developers of \Code{halo} and \Code{netcdf4} proposed to pin their versions to the version before the license change.

In three cases, developers conclude that remediation is not necessary because the incompatible dependency is optional (\Code{apache-airflow}, \Code{cvxpy}) or the dependency provides a dual-licensing option (\Code{dvc}). 
In the case of \Code{pytest-pylint}, developers questioned the necessity of remediation, but the issue is still open and unresolved at the time of writing.

\begin{small}
\begin{summary-rq}
\textbf{Answers for RQ3:}
PyPI package developers show unfamiliarity and raise controversies with OSS licensing when they discover a license incompatibility.
They remediate license incompatibilities by 1) migrating, removing, or pinning a version of the incompatible dependency; 2) changing their own licenses; or 3) asking upstream developers to change the licensing of their package. 

\vspace{-4pt}
\noindent\rule{\textwidth}{0.8pt}

\textbf{Implications:} 
Automated approaches can be helpful in making developers aware of license incompatibilities and recommending remediations. 
The practices taken by developers can serve as the solution space to be explored by automated approaches.

\end{summary-rq}
\end{small}

\section{The \ToolName~Approach}

Inspired by the results from the empirical study, we propose \ToolName, an \underline{S}MT-solver-based \underline{i}ncompatibility remediator for \underline{l}ic\underline{ense}s in the dependency graph.
In this section, we describe the design, implementation, and evaluation of \ToolName.

\subsection{Data and Notations}
\label{sec:notations}

Recall in Section~\ref{sec:def} that our dataset contains 438,967 packages and 3,622,711 releases from the entire PyPI ecosystem.
To simplify the presentation of \ToolName, we provide a formal notation of this dataset.
We denote the set of package names as $\mathcal{P}$, the set of version strings as $\mathcal{V}$, and the releases in our dataset (i.e., the entire PyPI ecosystem) as $\mathcal{E} \subseteq \mathcal{P} \times \mathcal{V}$ ($|\mathcal{E}| = 3,622,711$).
Each $\langle p, v \rangle \in \mathcal{E}$ contains:
\begin{itemize}
    \item An SPDX license identifier $l(p, v)$.
    \item Direct dependencies and version constraints $deps(p, v) \subseteq \mathcal{P} \times \mathcal{C}$ ($\mathcal{C} \subseteq \mathcal{V}^*$ denotes the set of version constraints).
    \item A dependency graph $\mathcal{G}(p, v) ::= \langle N(p, v), D(p, v) \rangle$, s.t. $\langle p, v \rangle \in N(p, v) \subseteq \mathcal{E}$, $D(p, v) \in N(p, v) \mapsto N^*(p, v)$.
    \item A list of incompatible dependencies $incomp(p, v) \subseteq N(p, v)$ in the dependency graph, such that $\langle p', v' \rangle \in incomp(p, v) \Rightarrow \langle l(p', v'), l(p,v) \rangle \in \mathcal{I}$ (here $\mathcal{I}$ denotes the set of one-way incompatible license pairs). 
\end{itemize}
We denote the set of 63 licenses in the compatibility matrix as $\mathcal{L}$.
To support finding migrations, we use the Python package migration dataset by Gu et al.~\cite{guMigration} containing 640 migration rules between Python packages, denoted as $\mathcal{M} \subseteq \mathcal{P} \times \mathcal{P}$.

\subsection{Problem Formulation}

According to our \textbf{RQ3}, developers may take one of the following approaches to remediate license incompatibilities: migration, removal, pinning version, changing their own license, and negotiating with upstream packages.
The results inspire us with the idea of using an automated approach to generate and recommend possible remediations to developers when a license incompatibility is detected (the detection can be easily automated using our PyPI dependency and licensing dataset in Section~\ref{sec:dataset}).
Such an automated approach can be implemented as a GitHub CI/CD Action or a bot deployed to notify and help developers remediate licensing incompatibilities.
As developers frequently discuss several remediations in their issues and choose one of them eventually, this automated approach should be able to recommend multiple reasonable remediations for developers to consider and choose. 

For the possible remediations, negotiations fall out of scope for an automation tool, and determining which license(s) can be changed is trivial as it only requires an enumeration of all alternative licenses while assessing their compatibility with the package dependency graph.
However, finding migration, removal, and version-pinning solutions is more challenging because incompatible dependencies may reside in a sophisticated dependency graph position and any change can have a ripple effect over the entire graph.
On the other hand, developers generally want to minimize changes to their dependency graph because larger changes would often result in more remediation effort.
What's more, finding viable migration targets itself is challenging and has been explored in prior research~\cite{DBLP:conf/wcre/HeXMXLZ21, DBLP:conf/icse/HeXCLZ21}.

Considering the above rationales, we define the license incompatibility remediation problem as follows:
\begin{enumerate}[leftmargin=12pt]
    \item \textbf{Input:} a release $\langle p, v\rangle$, its dependency graph $\mathcal{G}(p, v)$, and the PyPI dataset (Section~\ref{sec:notations});
    \item \textbf{Output:} $N$ alternative dependency graphs $\mathcal{G}'_1,...,\mathcal{G}'_N$, all of which have no license incompatibility and minimal changes to $\mathcal{G}(p, v)$, and $M$ alternative licenses $l_1,...,l_M$ with which $\langle p, v\rangle$ would have no license incompatibility in $\mathcal{G}(p, v)$.
\end{enumerate}

We observe that this definition is similar to the dependency resolution problem studied in prior work~\cite{DBLP:conf/kbse/MancinelliBCVDLT06, DBLP:journals/corr/abs-2203-13737, DBLP:journals/corr/abs-2301-08434} with some important differences.
The alternative dependency graphs can ignore dependencies (for removals), violate version constraints (for pinning versions), and add new direct dependencies (for migrations).
Nonetheless, any deviations from the original graph need to be minimized.
Just like the dependency resolution problem, such alternative dependency graphs can be found using a Max-SMT solver with a carefully designed objective function.
The exact remediations can be generated by comparing the alternative graph and the original graph.

\subsection{Approach Overview}

\vspace{-3mm}
\begin{algorithm}
\small
\caption{The \ToolName~Approach}
\label{alg:remediation}
\DontPrintSemicolon
\SetKwInput{Input}{Input}
\SetKwInput{Output}{Output}
\Input{$\langle p, v\rangle$, $\mathcal{G}(p, v)$, and the PyPI dataset (Section~\ref{sec:notations})}
\Output{$\mathbf{G} = \{\mathcal{G}'_1,...,\mathcal{G}'_N\}, \mathbf{L} = \{l_1,...,l_M\}$}
$\mathbf{G} \leftarrow \emptyset$, $\mathbf{L} \leftarrow \emptyset$\\
\ForEach(\Comment*[h]{find compatible licenses}){$l \in \mathcal{L}$}{
  \If{$\langle l(p',v'), l \rangle \notin \mathcal{I}$ for all $\langle p', v' \rangle \in N(p, v) \setminus \langle p, v\rangle$}{
    $\mathbf{L} \leftarrow \mathbf{L} \cup \{l\}$
  }
}
Keep only top-$M$ licenses in $\mathbf{L}$ ordered by their popularity\\
$vars \leftarrow \{p$, plus all packages reachable from $deps(p, v)\}$\\
$clauses \leftarrow \text{\textit{build\_constraints}}(p, v, vars)$\\
\While{$\mathcal{G}' \leftarrow \text{find\_solution}(vars, clauses, objective)$}{
  \textbf{if} $|\mathbf{G}|\ge N$ or $\mathcal{G}' = unsat$  \textbf{then break} \\
  $\mathbf{G} \leftarrow \mathbf{G} \cup \{\mathcal{G}'\}$\\
  Add new constraints to exclude solutions similar to $\mathcal{G}'$
}
\Return{$\mathbf{G}, \mathbf{L}$}
\end{algorithm}
\vspace{-3mm}

Algorithm~\ref{alg:remediation} summarizes the \ToolName~approach.
In line 2-5, it finds $M$ compatible licenses.
In line 6-10, it finds $N$ alternative dependency graphs without license incompatibilities.
The key idea is to find all packages that may be present in the alternative graph (line 6), build version constraint clauses for each package (line 7), and find top-$N$ solutions under the $objective$ function using a Max-SMT solver (line 8-11).
We will describe the underlying details in Section~\ref{sec:smt-solver}.

\subsection{SMT-Solver-Based License Incompatibility Remediation}
\label{sec:smt-solver}

To create a constraint SMT problem over a finite domain, the first step is to initialize a set of finite domain variables for all packages that may be present in the alternative graphs (i.e., $vars$ in line 6).
To find these packages, we utilize a breadth-first search (BFS) beginning from the root package $p$ and all possible migration targets $p_m$ that may replace one of the dependencies of $p$ (i.e.,  $\exists \langle p_d, C \rangle \in deps(p, v)$, s.t. $\langle p_d, p_m \rangle \in \mathcal{M}$).
For each package $p'$ in the BFS queue, we encode all its versions $v_1, ..., v_k$ in a finite integer domain, ordered by semantic versioning~\cite{SemanticVersioning}, from $-k$ to $-1$ (i.e., oldest to latest).
We use the special value $p'=0$ to indicate $p'$ is not included in the graph.
All packages that $p'$ may depend on (i.e., packages in $\bigcup_{i=1,...,k}deps(p',v_i)$) will be added to the BFS queue.
The search stops once saturation is reached (i.e., no more new packages could be added to $vars$).

With a set of finite domain variables $vars$, the next step is to encode their dependency relationships and version constraints as $clauses$ (line 7).
For each $p' \in vars$, excluding the root package $p$, we encode a logical implication as follows:
$$
(p'=v') \implies \bigwedge_{\langle p_d, C \rangle \in deps(p', v')}\left(\bigvee_{v_d \in C}\left(p_d=v_d\right)\right)
$$
Here we use $p = v$ as a convenience notation meaning that the corresponding finite domain variable of $p$ in $vars$ takes the concrete integer value corresponding to $v$.

For root package $p$, we need to encode possible remediations (i.e., migrations, removals, and version pinning) into its clause, all of which can result in violations of $dep(p, v)$.
To consider this, we add all possible migration targets without version constraints (i.e., $\{\langle p_m, \mathcal{V} \rangle\}$) to $deps(p, v)$, allow each $p_d$ in $deps(p, v)$ to be removed (i.e., $p_d=0$), and allow the version constraints to be violated, forming the following clause:
$$
    \bigwedge_{\langle p_d, C \rangle \in deps(p, v) \cup \{\langle p_m, \mathcal{V} \rangle\}}\left((p_d = 0) \vee \bigvee_{\langle p_d, v_d \rangle \in \mathcal{E}}(p_d=v_d)\right)
$$
Of course, $p$ must be of its original version (i.e., $p=v$).

Finally, to remediate license incompatibilities, we add the following logical implications for all packages in $vars$:

$$
\bigwedge_{p'\in vars}\left( \langle l(p', v'), l(p, v) \rangle \in \mathcal{I} \implies p' \ne v' \right)
$$

All the above $clauses$ form the constraints of this problem.
For the packages in $vars$, any set of concrete integer values satisfying all the constraints forms a valid solution.
However, the constraints here are loose with many possible solutions.
To find solutions (i.e., an alternative graph $\mathcal{G}'$) with minimal differences compared with the original graph $\mathcal{G}(p, v)$, we define the optimization $objective$ (in line 8) as follows:
$$
\min_{\mathcal{G}'}\sum_{\langle p_{\text{old}}, p_{\text{new}} \rangle \in \text{diff}(\mathcal{G}, \mathcal{G}')}
\left\{\begin{array}{lr}
c_{\text{migration}}, & \langle p_{\text{old}}, p_{\text{new}} \rangle \in \mathcal{M}\\
c_{\text{removal}}, & p_{\text{new}} = 0\\
|p_{\text{new}} - p_{\text{old}}|, & \text{otherwise}
\end{array}\right.
$$
Specifically, this objective function attempts to find a $\mathcal{G'}$ that minimizes the total cost of all changed packages by comparing  $\mathcal{G}'$ with $\mathcal{G}(p, v)$.
For each changed package, the cost depends on what has been changed: if there is a migration between two packages, we add a constant cost $c_{\text{migration}}$; if a package is removed, we add a larger constant cost $c_{\text{removal}}$; if the version is changed within the same package, we add a cost equal to the distance between the changed versions (i.e., $|p_{new} - p_{old}|$).
The two constant costs can be adjusted in practice.
Using this objective function, line 8-11 finds the top-$N$ solutions (ordered by the cost determined by $objective$) as the alternative graphs, all of which do not contain license incompatibilities.
To avoid generating redundant solutions, we add a new constraint to $clauses$ to exclude all solutions similar to the current solution:
$$
\bigvee_{\langle p_{\text{old}}, p_{\text{new}} \rangle \in \text{diff}(\mathcal{G}, \mathcal{G}')}
\left\{\begin{array}{lr}
p_{\text{new}} = 0, & p_{\texttt{new}} \ne 0\\
0, & \text{otherwise}
\end{array}\right.
$$
This means the new solution must not include all the changed packages in the previous solution.
The algorithm stops if the solver returns $unsat$ or it has found $N$ viable solutions.

\subsection{Implementation}
\label{Implementation}

We implement \ToolName~in Python using the Python binding of Z3~\cite{DBLP:conf/tacas/MouraB08}, the state-of-the-art SMT solver.
To find the versions satisfying version constraints, we 
simply use the Python standard library \Code{packaging} which implements PEP 440~\cite{PEP440}.
We also implement an additional post-processing step to convert the results of Algorithm~\ref{alg:remediation} into a remediation report like:
\begin{Report}
Possible Remediations for [package] [version]:
1. Change project license to $l_1$, $l_2$, ..., or $l_M$;
2. ($\mathcal{G}'_1$) Migrate [package] to [package];
3. ($\mathcal{G}'_2$) Remove [package];
4. ($\mathcal{G}'_3$) Pin [package] to [version];
\end{Report}

In current implementation, we heuristically set $N = 5$, $M = 3$, $c_{\text{migration}} = 10$, and $c_{\text{removal}} = 100$, which we find to produce satisfactory results (see Section~\ref{sec:evaluation}).
We tested \ToolName~on the 202 incompatible releases in \textsc{Top} (Table~\ref{tab:incom}). \ToolName~can generate results for all of them with a median running time of 14.9 seconds (max = 295 seconds), which is satisfactory in practical application scenarios (e.g., as a CI/CD workflow).

\subsection{Evaluation}
\label{sec:evaluation}

We evaluate the effectiveness of \ToolName~by observing to what extent can the remediations provided by \ToolName~match those \textit{proposed} by developers in the upper half of Table~\ref{tab:remediation}.
We do not compare against the final remediations because \ToolName~is intended to provide recommendations and support the decision-making process.
We observe that the final remediation is contingent upon multiple factors from the specific project context (e.g., the development cost of each remediation), so the decision should be left to project developers.

Of the 21 cases in upper Table~\ref{tab:remediation}, developers proposed at least one remediation in 19 cases, except for \Code{pytest-pylint} and \Code{dvc}.
We find that the results returned by \ToolName~can cover all the proposed removals, version-pinnings, and license change remediations in these cases.
However, due to the incompleteness of the Python migration dataset~\cite{guMigration}, \ToolName~can only cover two out of the 13 migration proposals in these cases (\Code{mysqlclient} to \Code{PyMySQL} and \Code{unidecode} to \Code{text-unidecode}).
For the remaining 11 migration proposals, \ToolName~simply proposes to remove the incompatible dependency, leaving developers to find migrations themselves.
This limitation can be easily overcome by adding more migration rules to $\mathcal{M}$ once they are discovered.
Based on this evaluation, we conclude that \ToolName~performs relatively well in the remediation of license incompatibilities for Python packages.

\subsection{Example}

In this section, we use the example of \Code{fiftyone 0.18.0} in Figure \ref{fig:example} to illustrate how \ToolName~can be applied to practice.
For \Code{fiftyone 0.18.0}, \ToolName~provides the following remediation report for the existing license incompatibilities:
\begin{Report}
Possible Remediations for fiftyone 0.18.0:
1. Change project license to GPL-3.0-only, 
    GPL-3.0-or-later, or AGPL-3.0-only;
2. Or make the following dependency changes:
    a) Remove ndjson;
    b) Pin voxel51-eta to 0.1.9;
    c) Pin pillow to 6.2.2;
    d) Pin imageio to 2.9.0;
    e) Pin h11 to 0.11.0.
3. Or make the following dependency changes:
    a) Remove voxel51-eta;
    b) Remove ndjson;
    c) Pin h11 to 0.11.0.
\end{Report}
This report includes changes to \Code{pillow} and \Code{imageio} due to the ripple effect of pinning \Code{voxel51-eta}.
The change to \Code{h11} is included to fix dependency conflicts in the previously resolved dependency tree, a positive side effect similar to SMT-solver-based dependency resolution like \textsc{smartPip}~\cite{DBLP:conf/kbse/WangWSSL22}.

As shown in Table~\ref{tab:remediation}, the developers of \Code{fiftyone} finally migrate \Code{ndjson}
to \Code{jsonlines}. 
As mentioned in Section~\ref{sec:evaluation}, this migration is not covered by an existing dataset~\cite{guMigration}.
By adding $\langle \Code{ndjson}, \Code{jsonlines}\rangle$ to $\mathcal{M}$, \ToolName~returns:
\begin{Report}
Possible Remediations for fiftyone 0.18.0:
1. Change project license to GPL-3.0-only, 
    GPL-3.0-or-later, or AGPL-3.0-only;
2. Or make the following dependency changes:
    a) Migrate ndjson to jsonlines;
    b) Pin voxel51-eta version to 0.1.9;
    c) Pin pillow to 6.2.2;
    d) Pin h11 to 0.11.0;
    e) Pin imageio to 2.9.0.
3. Or make the following dependency changes:
    a) Migrate ndjson to jsonlines;
    b) Remove voxel51-eta;
    c) Pin h11 to 0.11.0.
\end{Report}
With the above report, developers may conclude that \Code{ndjson} should be migrated to \Code{jsonlines}.
Although the report points out that removal or downgrading \Code{voxel51-eta} is necessary for remediating \Code{patool}, developers may find such remediation undesirable because \Code{voxel51-eta} is tightly integrated with \Code{fiftyone}.
In fact, they are developed under the same GitHub organization \Code{voxel51}.
In such cases, \textit{the dependency changes must be made upstream}.
The developers of \Code{fiftyone} may then begin to negotiate with the developers of \Code{voxel51-eta}, who can use \ToolName~to produce a report for themselves:
\begin{Report}
Possible Remediations for voxel51-eta 0.8.1:
1. Change project license to GPL-3.0-only, 
    GPL-3.0-or-later, or AGPL-3.0-only;
2. Or make the following dependency changes:
   a) Migrate ndjson to jsonlines;
   b) Migrate patool to py7zr.
3. Or make the following dependency changes:
   a) Migrate ndjson to jsonlines;
   b) Migrate patool to rarfile.
... (omitted due to space limitations)
\end{Report}

\subsection{Preliminary User Study}
\label{sec:user-study}

To evaluate how developers perceive the usefulness of \ToolName, we carefully select packages from \textsc{Top} that: 
1) have incompatible releases in our dataset; 
2) still have incompatibilities in their latest releases; 
3) actively use an issue tracker; 
4) have no previous issues about licensing. 
This results in ten packages.
After manual inspection, we exclude one false positive, \Code{dvc}, which is not actually incompatible with its GPL-licensed dependency \Code{pygit2} due to its explicit statement of link exception~\cite{PyGit2}.  
We then open nine issues with the report by \ToolName, summarized in the lower half of Table~\ref{tab:remediation}.

At the time of writing (August 2023), we received responses in seven issues, among which five packages have completely or partially adopted one of the remediations suggested by \ToolName. Notably, \Code{glean-parser} subsequently implemented license checking in its CI/CD workflow (\href{https://github.com/mozilla/glean_parser/pull/578}{\#578}), indicating the need for and usefulness of integrating tools like \ToolName~into the development process.
\Code{sphinx-autoapi} accepted the migration suggestion but migrated to another package not recommended by \ToolName. 
The remaining two packages, however, closed our issue.
One package responded that although they acknowledge this incompatibility, they will only fix it if it actually causes issues to end users (which they believe is unlikely because their package is a CI tool, not a library).

In conclusion, five of the seven responded packages adopted one of the suggestions provided by \ToolName. 
The high adoption rate signifies the relevance of license incompatibilities to PyPI developers, their positive attitude towards \ToolName, 
and the effectiveness of \ToolName~in addressing incompatibilities.

\section{Discussion}

\subsection{Implications}

In this section, we discuss the implications of our results for developers, package distribution platforms, and researchers.

\subsubsection{Developers} 

The results of \textbf{RQ1} show the prevalence of packages without accurate or complete licensing information in the PyPI ecosystem. 
However, if a package lacks licensing information, it is not really open-source~\cite{OSIdef}, posing difficulties for others to legally use this package. 
Hence, developers should pay meticulous attention to the licensing of their dependencies and provide precise licensing information for their own packages to the best of their abilities.
Additionally, 10.96\% of the \textsc{Top} packages have undergone at least one licensing change as revealed in \textbf{RQ1}, which may impact numerous downstream projects and lead to incompatibilities. Therefore, developers of popular and influential packages should exercise more caution than those of common projects when making decisions regarding licensing changes. 
Finally, RQ2 reveals that most of the license incompatibilities in the PyPI ecosystem are caused by direct dependencies (74.0\%). 
These incompatibilities can be easily detected by parsing dependency manifest files and license checking can be integrated into CI/CD workflow, as evidenced by our preliminary user study. However, an accurate dependency graph like in this paper is needed to thoroughly detect license incompatibilities. 

\subsubsection{Package Distribution Platforms}
In Section~\ref{sec:pypi-licensing}, we find that the license information of a large number of PyPI packages on the PyPI platform is missing and the \Code{license} field in the metadata does not have a uniform format, leading to the difficulty of identifying package's license. Therefore, package management platforms can enhance their management in this aspect by providing standardized options and requiring developers to provide accurate license information when uploading packages. Moreover, 
the platform can also perform license compatibility checks periodically, e.g., during the package uploading process, to ensure that the uploaded packages are compliant with licensing requirements.

\subsubsection{Researchers}
Our study sheds light on further research regarding license incompatibility.  
First, migration is the most common license incompatibility remediation practice (RQ3). Therefore, researchers can explore more accurate package migration recommendation techniques and build more comprehensive package migration datasets to help developers make more informed decisions. 
Second, we find that the licensing information declared by package developers is noisy. Therefore, better license detection techniques can be developed to capture these packages' licensing information in the future. 
Finally, our study also lays a foundation for further research on the license incompatibility remediation practices and automated solutions in other packaging ecosystems like NPM. 
\subsection{Limitations}
\label{sec:limitation}

We discuss some notable limitations of our dataset, the empirical study, and the \ToolName~approach, as follows.

In terms of the PyPI dependency dataset, its main limitation is that the resolved dependency graphs at time $t$ may differ from the actual dependency graphs resolved by popular tools like \Code{pip} or \Code{Poetry} at the same time.
However, since each of them uses different resolution algorithms and may change its algorithms in new versions (e.g., \Code{pip} implements backtracking since version 20.3~\cite{pipBacktracking}), we believe accurate historical replication is impossible.
Compared with using \Code{pip install}, our custom solver is orders of magnitude faster and able to resolve dependency graphs at arbitrary time points.
Despite possible deviations, we believe this approach is the most suitable for such large-scale studies (e.g., a similar approach is also used by Liu et al.~\cite{DBLP:conf/icse/LiuCF00022} for studying security vulnerabilities in npm).

Several limitations pertain to the PyPI licensing data.
First, this dataset does not consider dual licensing, multi-licensing, or license exceptions used by some OSS~\cite{DBLP:journals/iepol/CominoM11, DBLP:conf/icse/VendomeVBPGP17}. Although our manual evaluation shows that they are rare in PyPI (Section~\ref{sec:licensing-data-collection}), they may occasionally introduce false incompatibilities in the dataset.
Future work is needed to take these corner cases into consideration.
Second, our study ignores in-code licenses, which may also have incompatibilities with the package-wide license~\cite{tosem2022Inconsistencies}.
However, studying such incompatibilities would require a different methodology and is out of the scope of our study.
Finally, none of the authors are law professionals and the dataset may contain inaccurate license incompatibilities.
To alleviate this threat, we have tried our best to base our study on some sort of ``joint consensus'' among OSS developers, as reflected by reliable sources of information (e.g., OSI, FSF, and prior research).
Even if some of the data are proven to be incorrect, we believe the methodology and the \ToolName~approach presented in this paper are general and can be easily adapted to any new compatibility criterion.

In terms of external validity, the dataset and its construction process are largely unique and designed for PyPI, a flourishing packaging ecosystem of great importance in many application domains (e.g., AI). 
However, future work is needed for other packaging ecosystems, as they have different dependency resolution behaviors~\cite{DBLP:conf/wcre/AbateCGZ20} and licensing data format.
The remediation practices in RQ3 are identified from a small number of popular Python packages, but we believe the general pattern should be applicable to proprietary Python projects and even projects in other ecosystems (future work is needed to validate our belief).
The \ToolName~approach is also general and can be extended to other packaging ecosystems by taking their unique dependency resolution behaviors into consideration~\cite{DBLP:journals/corr/abs-2203-13737}.

\section{Conclusion}

In this paper, we contribute 1) a PyPI dependency \& licensing dataset, 2) a large-scale study of license incompatibilities and their remediation practices in the PyPI, and 3) an SMT-solver-based remediation approach, \ToolName.
As packaging ecosystems are likely to grow more complex~\cite{DBLP:journals/ese/DecanMG19}, we believe our contributions form a valuable reference for those willing to improve the state of OSS licensing compliance in modern packaging ecosystems. In the future, we plan to integrate our license incompatibility detection and remediation tool into CI/CD tools, e.g., GitHub workflow. 

\section{Data Availability}

We provide a replication package at:
\begin{quote}
    \url{https://github.com/osslab-pku/SILENCE}
\end{quote}

\section*{Acknowledgment}
This work is sponsored by the National Natural Science Foundation of China 61825201. We would like to extend our appreciation to Chao Wang and Xin Wu for their invaluable insights on license compliance. 

\bibliographystyle{IEEEtran}
\bibliography{reference}

\end{document}